\documentclass[manuscript]{aastex} 

\shorttitle{HIGH-FREQUENCY GLOBAL WAVES IN THE SUN}
\shortauthors{Brajesh Kumar et al.}

\begin{document}

\title{ON THE FLARE INDUCED HIGH-FREQUENCY GLOBAL WAVES IN THE SUN}

\author{BRAJESH KUMAR}
\affil{Udaipur Solar Observatory, Physical Research Laboratory,
 Dewali, Badi Road, Udaipur 313 004, India}
\email{brajesh@prl.res.in}
\and
\author{SAVITA MATHUR}
\affil{Indian Institute of Astrophysics, Koramagala, Bangalore 560 034, India}
\email{smathur@iiap.res.in}
\and
\author{R. A. GARC\'IA}
\affil{Laboratoire AIM, CEA/DSM-CNRS, Universit\'e Paris 7 Diderot, IRFU/SAp, 
Centre de Saclay, 91191, Gif-sur-Yvette, France}
\email{rafael.garcia@cea.fr}
\and
\author{P. VENKATAKRISHNAN}
\affil{Udaipur Solar Observatory, Physical Research Laboratory, Dewali, Badi Road, 
Udaipur 313 004, India}
\email{pvk@prl.res.in}

\begin{abstract}

Recently, \cite{karoff08} presented evidence of strong correlation between the 
energy in the high-frequency part (5.3$<$$\nu$$<$8.3 mHz) of the acoustic spectrum of the Sun and 
the solar X-ray flux. They have used disk-integrated intensity observations of the Sun obtained from 
the VIRGO (Variability of solar IRradiance and Gravity Oscillations) instrument on 
board {\em SOHO (Solar and Heliospheric Observatory)} spacecraft. Similar 
signature of flares in velocity observations has not been confirmed till now. The study of 
low-degree high-frequency waves in the Sun is important for our understanding of the dynamics of 
the deeper solar layers. In this paper, we present the analysis of the velocity observations of 
the Sun obtained from the MDI (Michelson and Doppler Imager) and the GOLF (Global 
Oscillations at Low Frequencies) instruments on board {\em SOHO} for some major flare events of the 
solar cycle 23. Application of wavelet techniques to the time series of disk-integrated velocity 
signals from the solar surface using the full-disk Dopplergrams obtained from the MDI 
clearly indicates that there is enhancement of high-frequency global waves in the Sun during the flares. 
This signature of flares is also visible in the Fourier Power Spectrum of these velocity oscillations. 
On the other hand, the analysis of disk-integrated velocity observations obtained from the GOLF shows 
only marginal evidence of effects of flares on high-frequency oscillations. 
    
\end{abstract}

\keywords{Sun: atmosphere, Sun: flares, Sun: oscillations}

\section{INTRODUCTION}

In the Sun, the typical distribution of acoustic power of photospheric oscillations 
peaks at around 5 min \citep{leighton62} with a general decrease 
to negligible power at higher frequencies. This behaviour of the acoustic spectrum has been 
understood in terms of trapped oscillations in a cavity \citep{ulrich70, leib71}. 
The eigen-functions corresponding to the band of 
oscillations in the region of 5 minutes peak in the convection zone, which seems to be 
the dominant source of excitation of the solar oscillations (Goldreich, Murray, and Kumar, 1994). 
Apart from these normal modes of oscillations ($p$ modes), 
researchers have found the presence of high-frequency oscillations (frequencies higher 
than the solar-photospheric acoustic cutoff at about 5.3 mHz) in the solar-acoustic spectrum 
\citep{libb88a, libb88b, chap03, jim05, karoff08}. Unlike 
the normal $p$ modes, the driving mechanism of these high-frequency solar oscillations 
is still not well established. \cite{balm90} suggest 
that the high-frequency waves are partly reflected by the sudden change in temperature at the transition
region between the chromosphere and the corona. \cite{kumar91} explain these high-frequency waves as 
an interference phenomenon between ingoing and outgoing waves from a localized source just beneath 
the photosphere. 

\cite{wolff72} suggested that large solar flares can stimulate free modes of oscillation of the 
entire Sun, by causing a thermal expansion that would drive a compression front to move 
into the solar interior. Several researchers have tried to study the effect of flares on the 
acoustic-velocity oscillations of the Sun. The progress in this 
field escalated recently with the advent of continuous data from dedicated instruments 
such as MDI (Michelson and Doppler
Imager) \citep{scherrer95} on board {\em SOHO (Solar and Heliospheric Observatory)} 
spacecraft and GONG 
(Global Oscillation Network Group) \citep{harvey95}. \cite{haber88} reported 
an average increase in the power of intermediate-degree modes after a major flare 
(of class X13/3B) using a few 
hours of solar-oscillations data obtained by the Dunn telescope at National Solar 
Observatory/ Sacramento Peak, USA. \cite{braun90} could not detect acoustic-wave excitation 
from an X-class flare. \cite{koso98} reported the first detection 
of ``solar quakes'' inside the Sun, caused by the X2.6 flare of 9 July 1996, using the MDI 
Dopplergrams.  Following this result, \cite{donea99} 
found an acoustic source associated with a flare using seismic images produced with 
helioseismic-holography technique. Application of ring-diagram analysis showed that the 
power of the global $p$ modes appears 
to be larger in several flare-producing active regions as compared with the power in 
non-flaring regions of similar magnetic field strength \citep{ambastha03}. 

Additionally, \cite{donea05} have reported emission of seismic waves from large 
solar flares using helioseismic holography. Some of the large solar flares have been observed to
produce enhanced high-frequency acoustic velocity oscillations in localized parts of active regions
\citep{kumar06}. Further, \cite{zharkova07} reported large downflows associated 
with seismic sources during the major flare (of class X17.6/4B) of 28 October 2003. 
\cite{venkat08} report co-spatial evolution of seismic sources and 
H-alpha flare kernels during the aforementioned flare event. A search for a correlation between 
the energy of the low-degree $p$ modes and flares using velocity observations 
of the Sun remained inconclusive \citep{gavry99, chap04, ambastha06}. The study of 
low-degree high-frequency (LDHF) waves in the Sun is important as this can bring more constraints 
on the rotation profile between 0.1 and 0.2~R$_{O}$ \citep{garcia_new08, mathur08}. The effect of flares 
on such LDHF waves can provide a clue for the origin of these waves.

Recently, \cite{karoff08} have reported that the correlation between X-ray flare 
intensity and the energy in the acoustic spectrum of disk-integrated intensity oscillations 
(as observed with VIRGO (Variability of Solar IRradiance and Gravity) \citep{frohlich95} 
instrument on board {\em SOHO}) is stronger for 
high-frequency waves than for the well known 5-minute oscillations. This correlation was obtained 
for a long time series of 8 years with time variation of power smoothed with a 24-day filter. 
This result does not show the immediate consequence of individual flares. For this, we require 
shorter time series closely associated with the flares. Since, the power spectra are noisier 
for shorter time series, we need to apply wavelet analysis to the time series. Intensity variations 
are generally manifestations of pressure variations. However, for a medium stratified by gravity, 
the waves are not pure acoustic waves, but are acousto-gravity waves. Hence, it will be 
more profitable to look at the velocity oscillations of the solar atmosphere. Therefore, in this 
study, we have searched 
for the effects of flares in the time series of disk-integrated velocity 
signals from the solar surface using the full-disk Dopplergrams obtained from the MDI instrument. 
We have also looked for these effects in disk-integrated velocity observations obtained from 
the GOLF (Global Oscillation at Low Frequency) 
\citep{gabriel95, gabriel97} instrument on board {\em SOHO}. These studies have been applied to the 
major solar flares of 28 October 2003 (of class X17.6/4B), 29 October 2003 (of class X10/2B) and 
6 April 2001 (of class X5.6/3B) that occurred in the solar cycle 23. Each of the selected events 
are in decreasing 
order of flare strength as observed in X-ray (1$-$8 \AA) by {\em GOES} (Geostationary Operational 
Environmental Satellite) \citep{garcia94} satellites. Wavelet and Fourier analyses of MDI 
velocity observations clearly indicate the enhancement in high-frequency global waves in the 
Sun during the flares. However, this signature of flares is weaker in the case of GOLF 
as compared to MDI data.

\section{DATA AND ANALYSIS}

\subsection{MDI DATA}

We have used the sequence
of MDI full-disk Dopplergrams for three hours spanning the flare obtained on 28 October 2003 
(10:00--13:00~UT), 29 October 2003 (19:00--22:00~UT), and 6 April 2001 (18:00--21:00~UT) with a 
cadence of one minute and spatial sampling rate of 2 arcsec per pixel. 
We have also used three hours of MDI 
Dopplergrams for a quiet period (non-flaring condition) as control data. In order to compare the 
temporal behaviour with the 
disk-integrated intensity observations by VIRGO, we have summed up the velocity signals over all 
the pixels of the MDI full-disk Dopplergrams. The 
images are first examined for pixels having cosmic ray hits and such pixels are replaced by 
interpolation. A two-point 
backward difference filter (difference between two consecutive measurements) 
is applied to these sequences of images to enhance the velocity signals 
from the $p$ modes and high-frequency waves above the solar background. The sequence of these filtered Doppler 
images are then collapsed into a single velocity value, excluding 
the noisy pixels along the solar limb. This process is applied to the time series of Doppler images 
for all the three aforementioned flare events as well as the quiet period. It is believed that by 
collapsing the 
full-disk Doppler images, the acoustic modes with maximum $l$=0,1,2,3 remain while the modes higher than 
these are averaged out. Thus, the collapsed velocity value should be the representative of the global 
acoustic modes. The temporal evolution of these disk-integrated velocity signals for three hours 
spanning the flare are shown in the upper panel of Figures~1(a), 2(a) and 3(a) and that for the 
quiet period is shown in Figure~4(a). 

\subsection{GOLF DATA}

GOLF instrument measures the disk-integrated line-of-sight velocity 
of the Sun at a cadence of one raw count every 10~s. However, we have rebinned the velocity data 
from GOLF for every 60 s to match with MDI. We have used the same periods of time for both instruments 
for the flare events of 
28 October 2003, 29 October 2003, and 6 April 2001, as well as for the non-flaring condition. We have 
worked with the standard velocity time 
series \citep{garcia05, ulrich00} as well as the original raw-counting rates. We have also checked if 
there was any 
anomalous behavior during the analyzed periods of time in the housekeeping data. Indeed, during the 
flare events of 28-29 October 2003, GOLF raw-counting rates of both photomultipliers suffered an increase 
in the measurements probably as a consequence of the impact of high-energetic particles. Due to these 
contaminations, the standard velocity time series have been filtered out during this period and thus, 
we have been obliged to work with the raw observations obtained by GOLF during the aforementioned 
flare events. A two-point backward difference 
filter is applied to the velocity series to remove the effect of the rotation and other slowly varying 
solar features. The temporal evolution of these filtered velocity signals for three 
hours spanning the flare are shown in the upper panels of Figures~1(b), 2(b) and 3(b) and that for 
the quiet period is shown in Figure~4(b).

\subsection{WAVELET ANALYSIS OF MDI AND GOLF VELOCITY DATA}

For the time-frequency analysis of the influence of flares on the high-frequency acoustic modes, we have 
applied the Wavelet technique \citep{torrence98} on the velocity time series obtained from MDI 
and GOLF as described above. We have used the Morlet wavelet, which is the product of a sine wave and a 
Gaussian function. We have computed the Wavelet Power Spectrum (WPS), which yields the correlation between 
the wavelet with a given frequency and the data along time. For the different events and the two instruments, 
we notice the presence of power around 3~mHz, corresponding to the region of the normal $p$ modes. 
Concerning the high-frequency waves above 5~mHz, we can study the evolution of their power with time. 
We limit our study to the region 
inside a ``cone of influence'' corresponding to the periods of less than 25\% of the time series 
length for reliability of the periods. Finally, we have also calculated two confidence levels of 
detection corresponding to the 
probability of 90\% and 50\% that the power is not due to noise. Thus, we have outlined the regions in the 
WPS where power lies above these confidence levels and these regions are shown in middle panels of the 
Figures~1-4(a)\&(b). A comparison of the WPS for the flare events with that of the quiet period 
clearly indicates the enhancement of high-frequency waves during these flare events as seen in the data 
from MDI instrument. However, in the case of GOLF data, some short-lived high-frequency waves are 
sporadically observed during the flares.
 
The WPS is collapsed along time to obtain the Global-wavelet power spectrum (GWPS). If some power is 
present during the whole length of our time series, it would be easily seen in the GWPS. This is 
nearly similar 
to the commonly used power spectral density. In the Figures~1-4(a)\&(b), the GWPS shows a strong 
peak of the normal 
$p$ modes which are well known to be existing all the time. However, if the power of high-frequency 
waves increases only a few times along the three hours of the studied data this would not appear as a 
strong peak in the GWPS. In case of MDI data, the GWPS for all the three flare events do show a bump 
corresponding to the high-frequency waves (above 5~mHz). Here again, we have overplotted the significance 
levels corresponding to probabilities of 90\% and 50\%. We notice that the normal $p$-mode peak 
in the GWPS is above 90\% significance level whereas the high-frequency peak is just below 50\% 
significance level, i.e., it has at least 50\% probability of being due to noise. However, the GWPS estimated 
from the MDI velocity data during a quiet period (Figure~4(a)) doesn't show any peak beyond 5~mHz. 
This supports the idea that the increase of power in the high frequency regime of the GWPS is 
indeed caused by the flare. In the case of GOLF data, we do not observe this 
signature in the GWPS estimated for the flare events as the overall high-frequency signal is weak in 
these observations.

To see when the high-frequency waves have an increased power during the flare, we have calculated the 
scale-average time series in the frequency regime 5-8~mHz. Basically, it is a collapsogram of the WPS 
along the frequency of the wavelet in the chosen range. For this quantity, we have calculated the 
confidence level for a 50\% probability. These are shown in the lower panels of the Figures~1-4(a)\&(b). 
Here, we observe peaks corresponding to the presence of high-frequency waves in the WPS. In general, the 
MDI data show more closely spaced high-amplitude peaks as compared to the GOLF data for the flare events, 
but still mainly around 50\% confidence level. A comparison 
with the same analysis performed for a quiet period (non-flaring condition) neither shows a high-frequency 
bump beyond 5~mHz in the GWPS (for MDI data) nor high-amplitude peaks in the scale-average time series 
(for both, MDI and GOLF data). Thus, in spite of the small confidence 
levels found during this analysis, it indicates a possible relationship between these excess 
high-frequency power and the flares.

\subsection{FOURIER ANALYSIS OF MDI AND GOLF VELOCITY DATA}

We have also estimated the Fourier Power Spectrum (FPS) from the velocity time series obtained by the 
MDI and GOLF instruments for the aforementioned flare events and the quiet period. The FPS spectra obtained 
from the MDI and GOLF data 
are respectively shown in the left and right panels of Figure~5. In the estimation of FPS spectra, 
the Power Spectral Density has 
not been corrected for the transfer function of the two-point backward difference 
filter \citep{garcia08}. A correction from the backward difference 
filter will change the slope, thereby affecting the relative amplitudes between the normal $p$ modes 
and the high-frequency waves. Here, we study the high-frequency waves in an absolute manner for each 
event and we do not compare them with the normal $p$ modes. Therefore, it does not bias our study. 
The FPS shown in the Figure~5 depicts dominant power in the 3~mHz band which 
is due to the normal $p$ modes. These FPS also show significant peaks in the higher frequency 
band (above 5~mHz) as estimated from MDI data for all the three flare events (strongest for the 
6 April 2001 flare event). However, the GOLF data show the strong
signature only for the flare event of 29 October 2003. The enhancement of high-frequency power related 
to the flare of 28 October 2003 has already been seen in the GONG velocity data \citep{kumar09}. 
Here, we do observe spikes at high-frequency 
in the FPS from the MDI data and some smaller ones from the GOLF data for these flare events. These spikes 
are very weak in the case of a quiet period, as seen in both the data sets (c.f., bottom panels of Figure~5).

The difference found between the measurements of the two instruments could be a direct consequence 
of the different heights of the solar atmosphere sampled by each instrument (Ni~I v/s Sodium doublet). 
The GOLF is observing in the high photosphere and the low chromosphere while the MDI or the GONG 
observes mostly in the deep photosphere close to where VIRGO is observing. Further analysis will be 
necessary to understand all the particularities of each seismic measurements (c.f. \cite{jim07}).

\section{DISCUSSION AND CONCLUSIONS}

The existence of high-frequency acoustic waves was first discovered in high-degree observations 
at the Big Bear Solar Observatory \citep{libb88a, libb88b} and 
later in GOLF low-degree disk-integrated observations \citep{garcia98}. Recently, they 
have also been seen in BiSON disk-integrated radial-velocity data \citep{chap03} and 
VIRGO intensity data \citep{jim05}. However, earlier attempts to find a correlation between 
the energy of these high-frequency oscillations and flares using disk-integrated velocity observations 
of the Sun had remained unclear \citep{gavry99, chap04}. In our analysis, the 
enhancement of high-frequency power is clearly seen in the 
MDI velocity data (and a feeble enhancement is also seen in the GOLF velocity data) during the 
major flare events. It is comparable with the flare related enhancements 
reported by \cite{karoff08} in disk-integrated intensity oscillations as observed with VIRGO. 
Although, in our results the flare induced enhancement signals are seen with a low probability (around 50\%), 
this signature is larger than that seen in non-flaring condition for both the MDI and GOLF data. 

Basically, two models have been proposed to account for these low-degree high-frequency oscillations with 
frequencies higher than the photospheric acoustic cut-off frequency ($\sim$5.3~mHz). The first model 
proposed by \cite{balm90} suggests that the high-frequency waves are partially reflected at the transition
region in comparison to the photospheric reflection for ordinary $p$ modes. The other model proposed 
by \cite{kumar91} explains the high-frequency waves as 
an interference phenomenon between ingoing and outgoing waves from a localized source just beneath 
the photosphere. In either case, the amount of energy that is stored in the high-frequency waves is extremely 
low compared to the amount of energy stored in the normal $p$ modes which are powered by the strong 
turbulence in the convection zone of the Sun. Therefore, it is believed that the flare energy will have 
a larger relative effect at high frequency as the other sources of its excitation are much smaller. 
These observations open a new area of study concerning the excitation of global high-frequency waves 
by local tremors due to major solar flares. It would be interesting to correlate the epochs of 
enhancement of high-frequency waves with episodes of flare energy release, using a similar analysis 
with hard X-ray data. We defer this study to the future. 

\acknowledgments
The use of data from the MDI and the GOLF instruments on board {\em SOHO} spacecraft is gratefully 
acknowledged. 
The {\em SOHO} is a joint mission under cooperative agreement between ESA and NASA. This work has 
been partially supported by the CNES/GOLF grant at the Service d'Astrophysique (CEA/Saclay). 
We are thankful to the referee for very useful suggestions which improved the presentation of this 
work. We are also thankful to Douglas Gough, 
John Leibacher, Frank Hill, P. Scherrer, Robertus Erdelyi, 
H. M. Antia, A. Kosovichev and Christoffer Karoff for useful discussions related to this work.


\begin{figure}
\centering
\plottwo{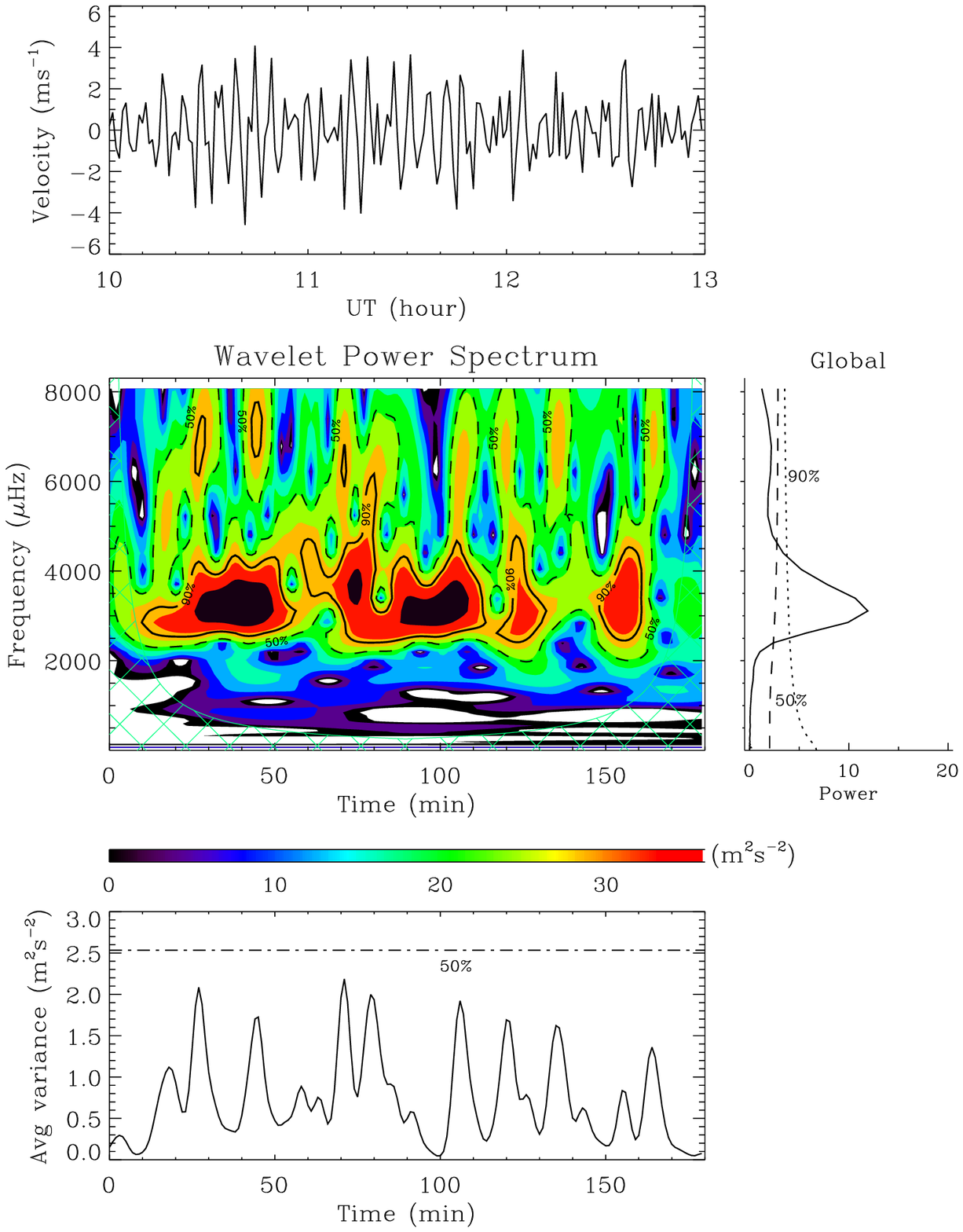}{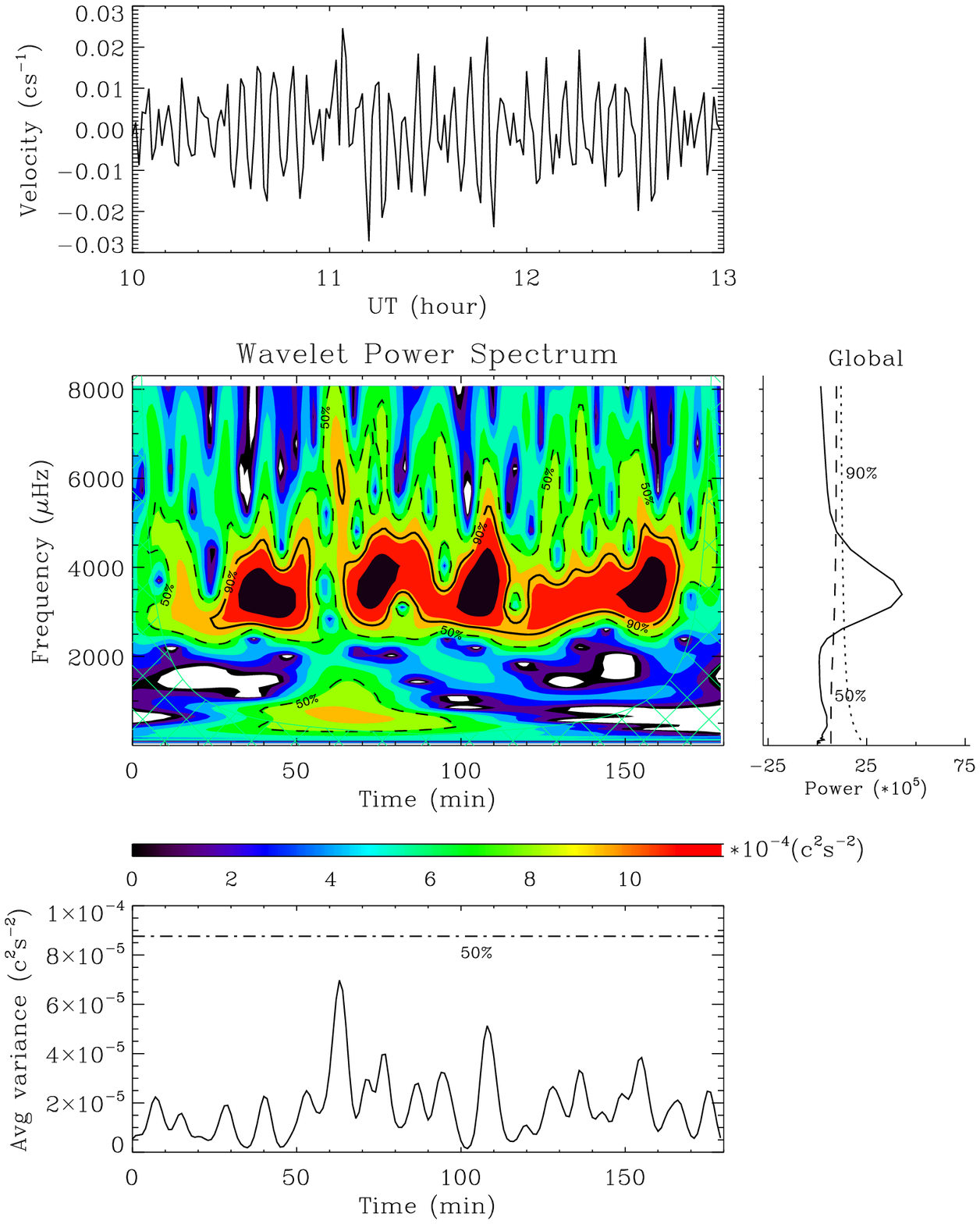}\\
\hspace*{0.7cm}(a)
\hspace*{7.0cm}(b)
\caption{(a) The upper panel shows the temporal evolution of disk-integrated velocity signals obtained from the 
full-disk Dopplergrams observed by MDI during 10:00-13:00~UT spanning the flare event of 28 October 2003. 
The middle panels show the Wavelet Power Spectrum (WPS) and the Global-Wavelet Power Spectrum (GWPS) computed 
from this time series. In the WPS, the solid lines correspond to regions with 90\% confidence level whereas 
the dashed lines are for 50\% confidence level and the hatched region indicates the cone of influence. The 
color scale is for the wavelet power. In the GWPS, the dotted line is for 90\% significance level 
and the dashed line is for 50\% significance level. The bottom panel illustrates the scale-average 
time series for the WPS in the frequency regime 5-8~mHz. The dashed-dotted line corresponds to 50\% 
significance level of the average variance. (b) Same as above, but using GOLF data.}
\end{figure}

\begin{figure}
\centering
\plottwo{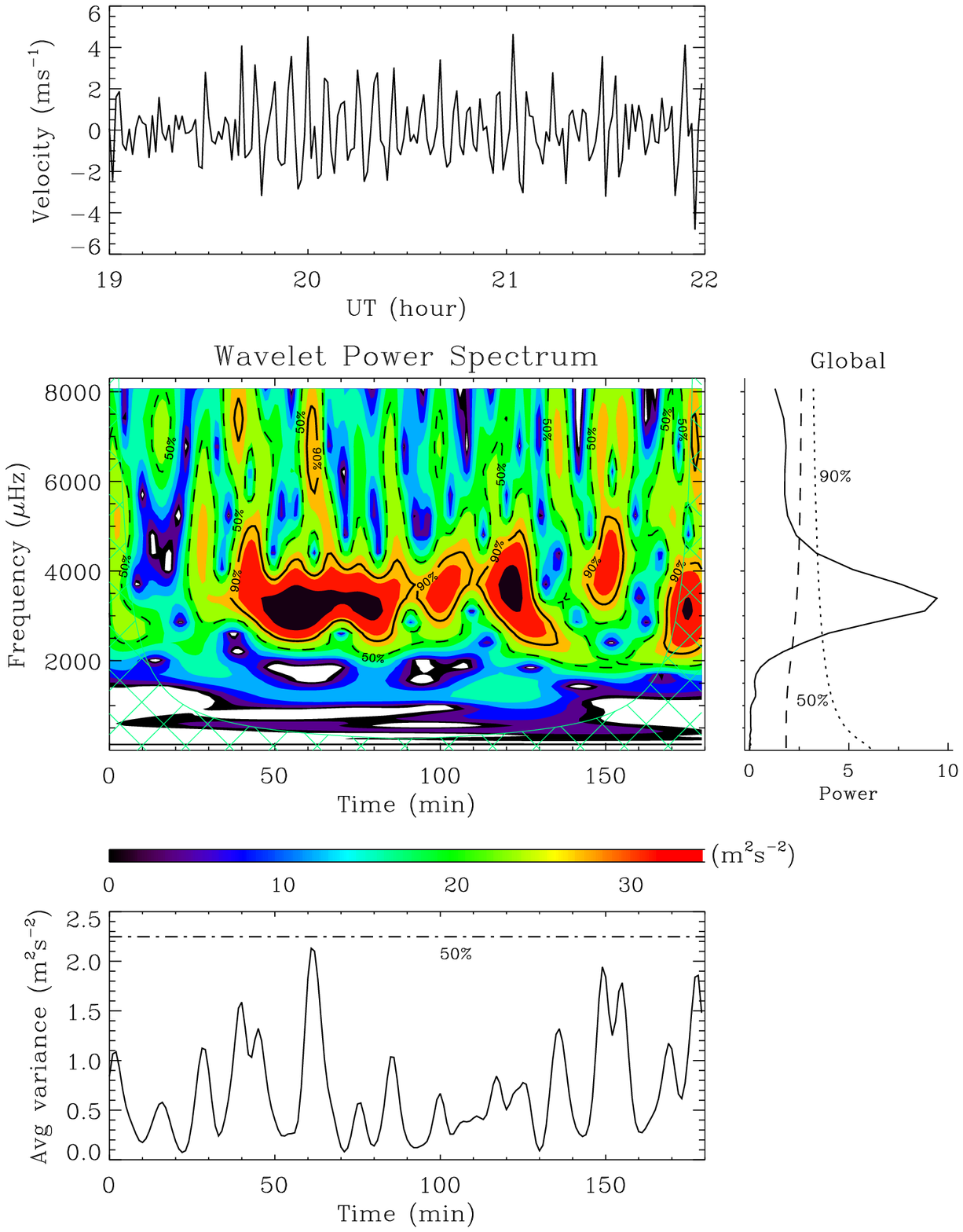}{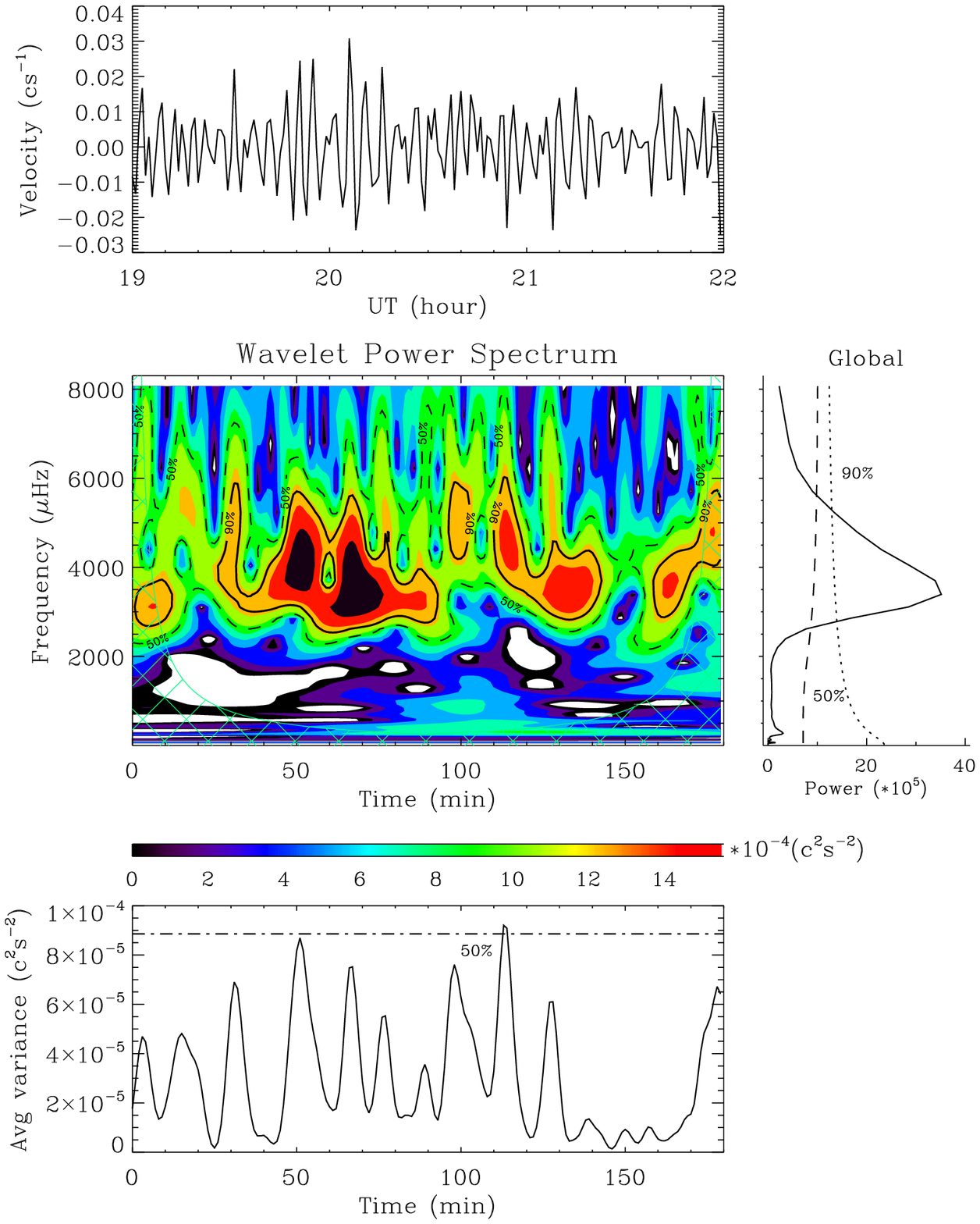}\\
\hspace*{0.7cm}(a)
\hspace*{7.0cm}(b)
\caption{Same as Figure~1, but for the flare event of 29 October 2003 during 19:00-22:00~UT using 
(a) MDI data and (b) GOLF data.}
\end{figure}

\begin{figure}
\centering
\plottwo{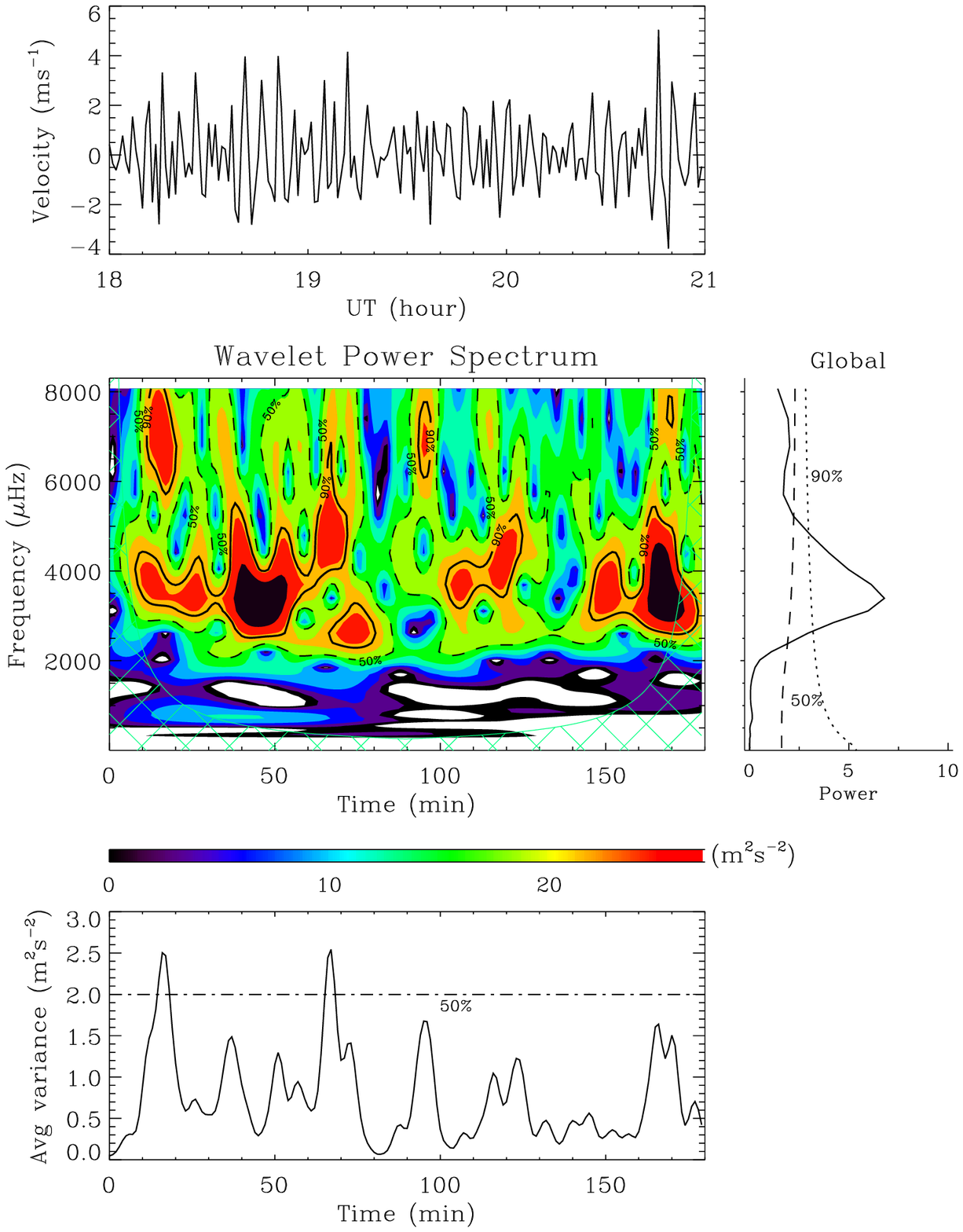}{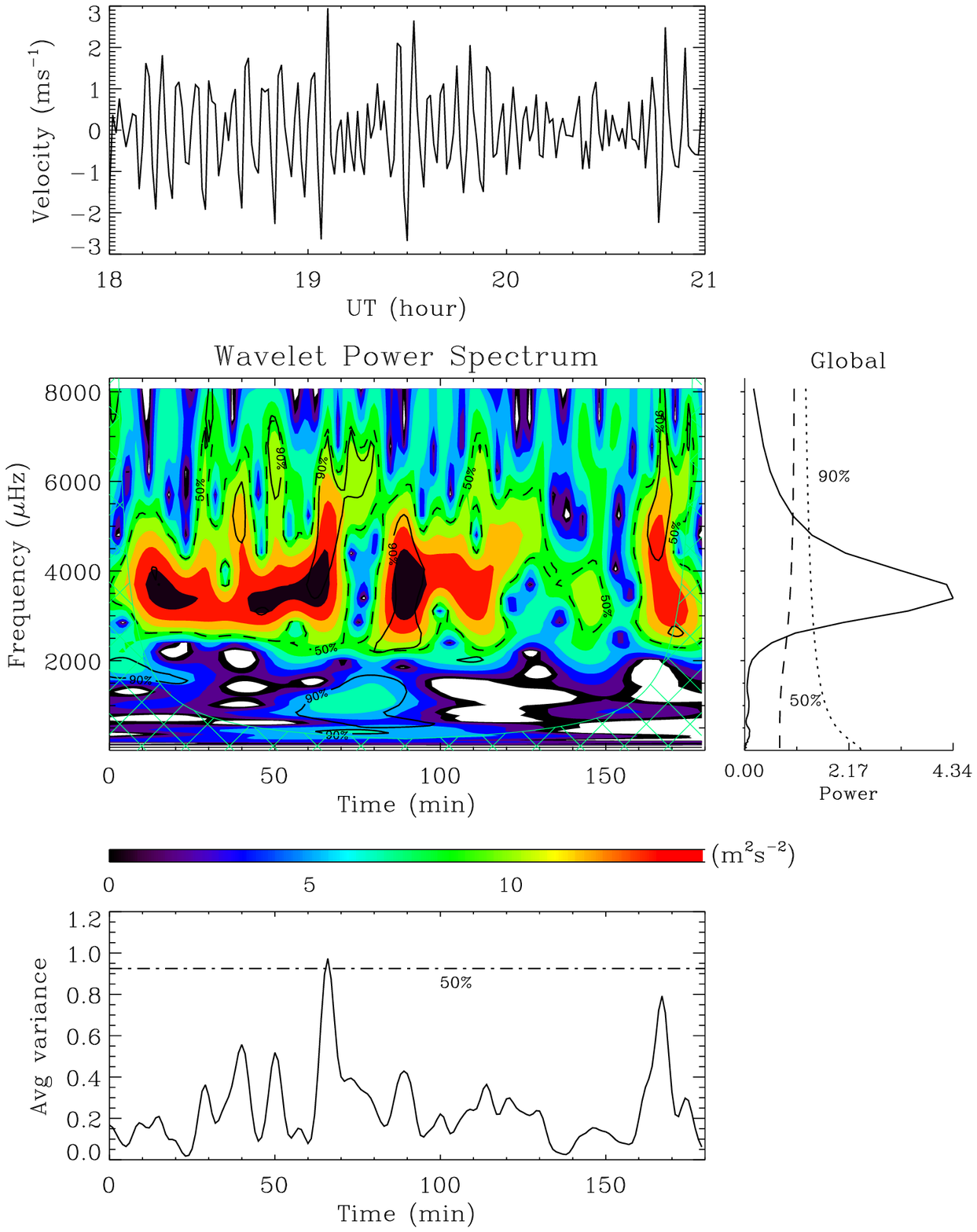}\\
\hspace*{0.7cm}(a)
\hspace*{7.0cm}(b)
\caption{Same as Figure~1, but for the flare event of 6 April 2001 during 18:00-21:00~UT using (a) MDI data 
and (b) GOLF data.}
\end{figure}

\begin{figure}
\centering
\plottwo{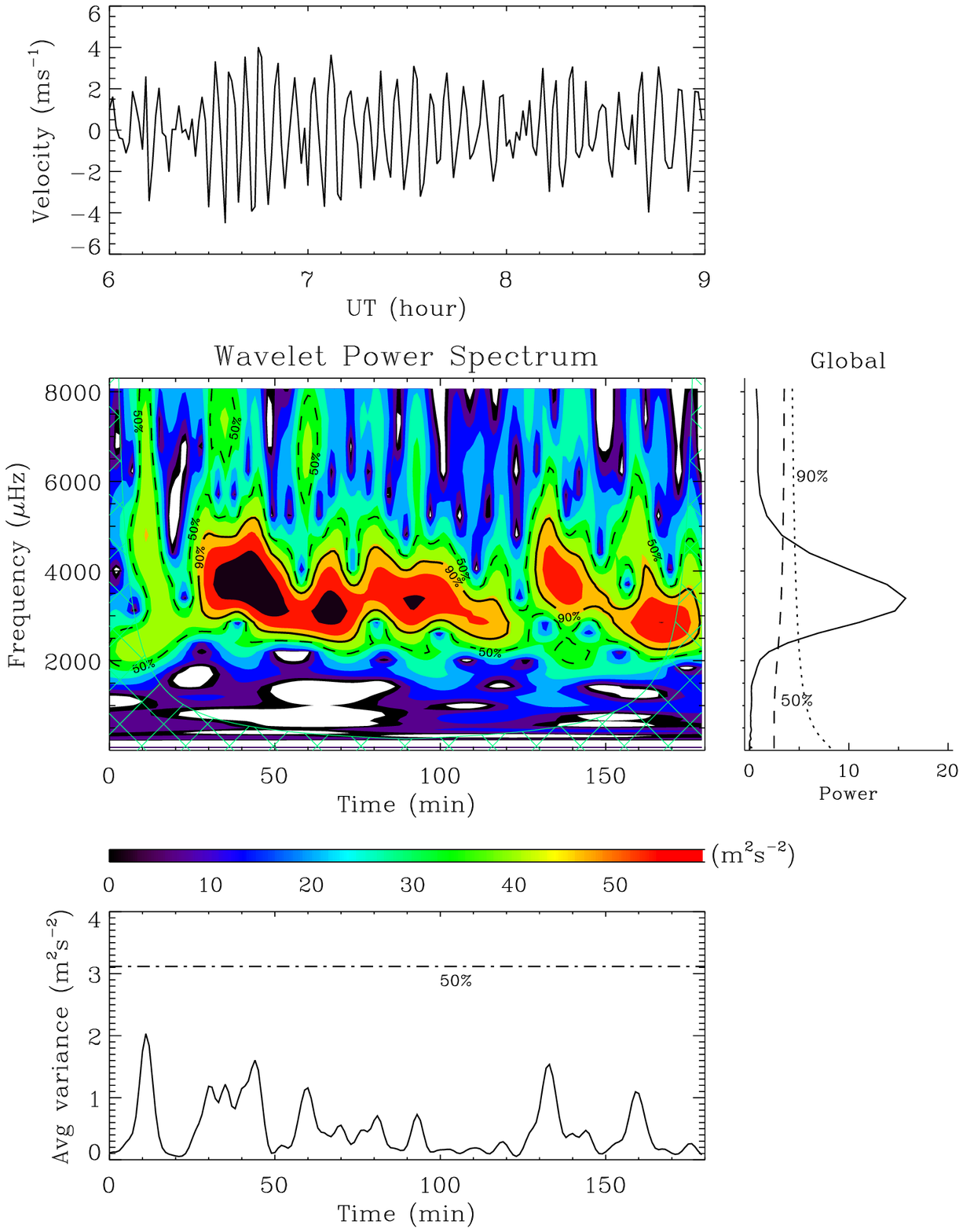}{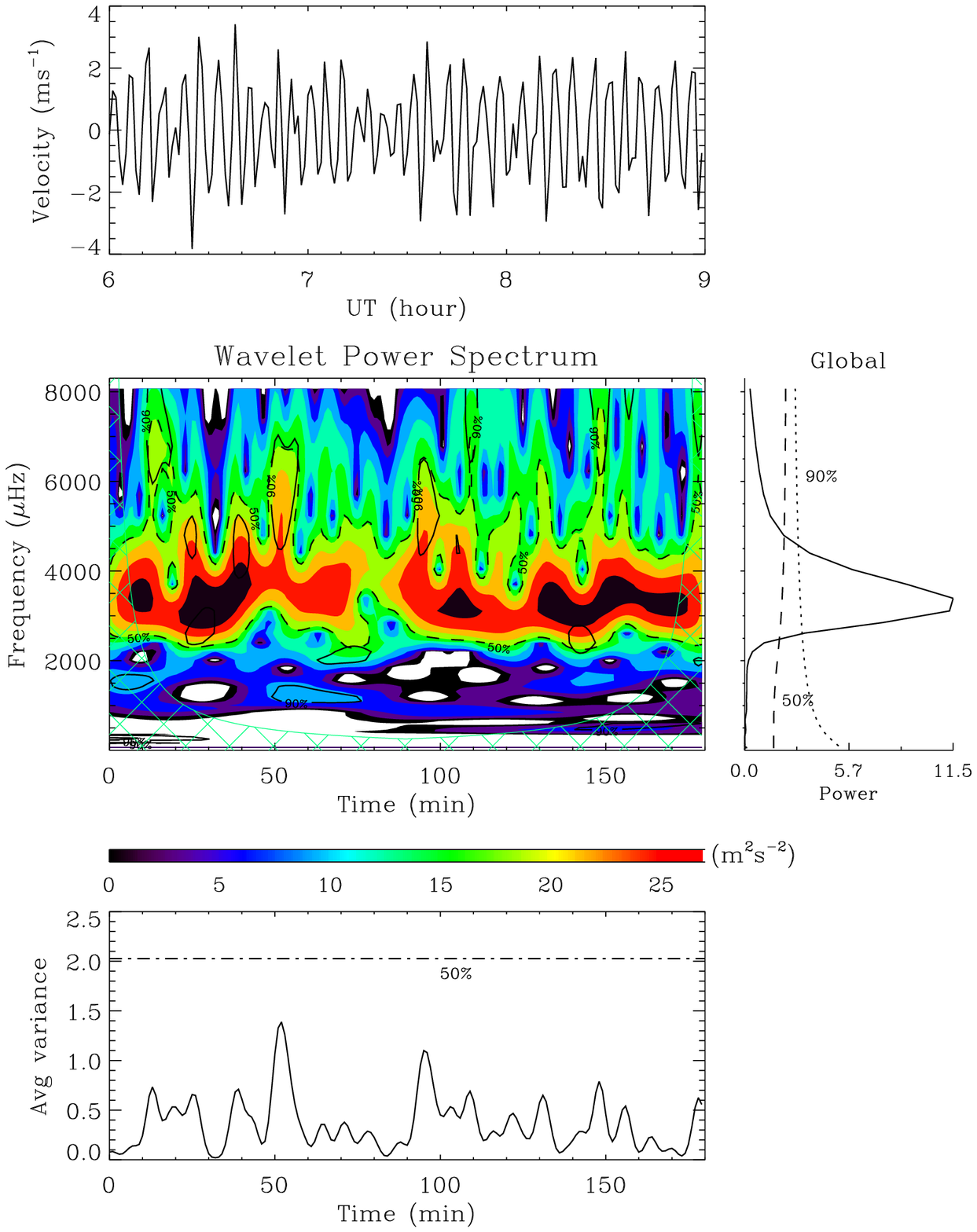}\\
\hspace*{0.7cm}(a)
\hspace*{7.0cm}(b)
\caption{Same as Figure~1, but for a quiet period (non-flaring condition) using (a) MDI data and (b) GOLF data.}
\end{figure}

\begin{figure}
\centering
\includegraphics[width=0.25\textwidth, angle=90]{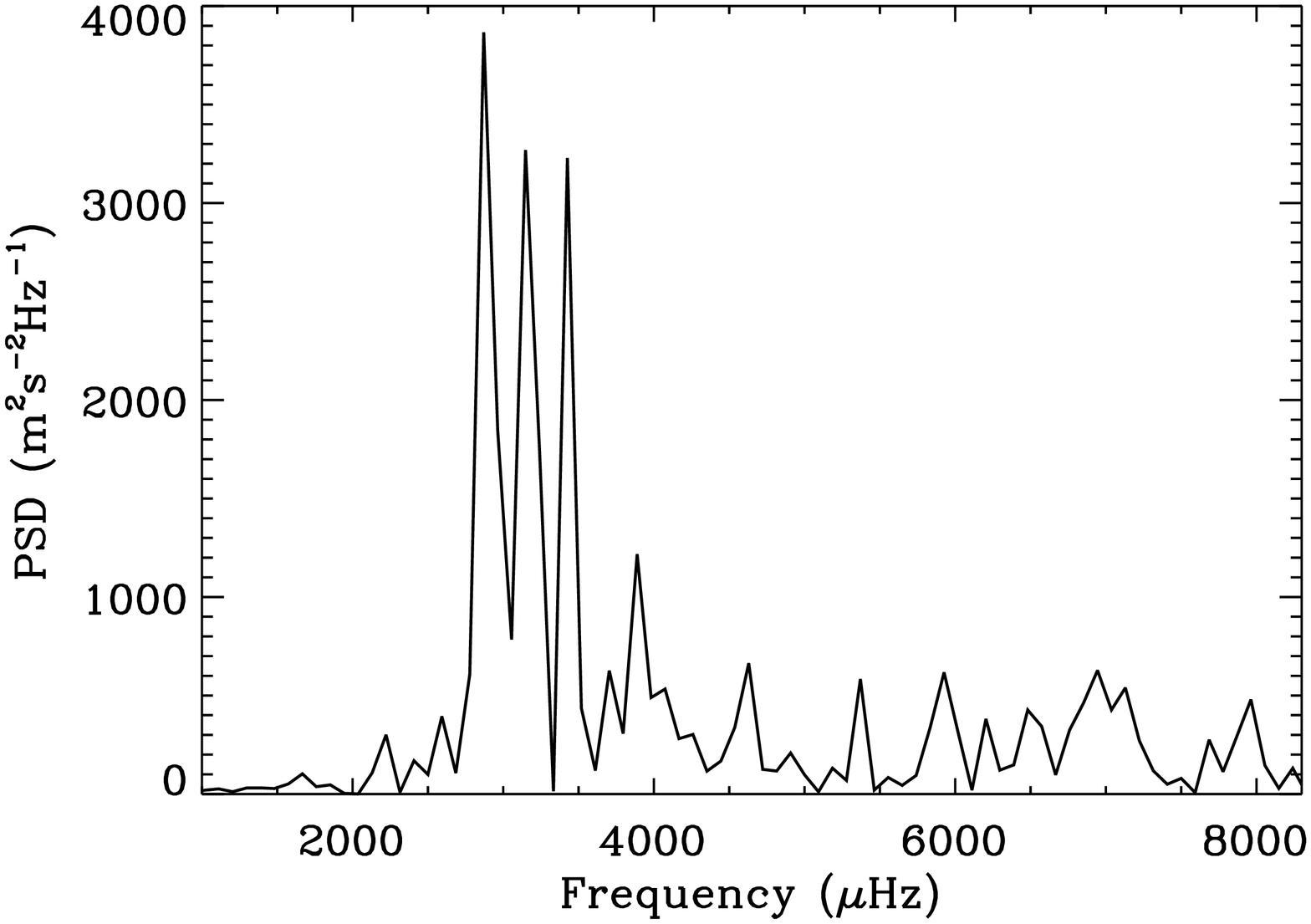} \hspace*{0.25 cm} 
\includegraphics[width=0.25\textwidth, angle=90]{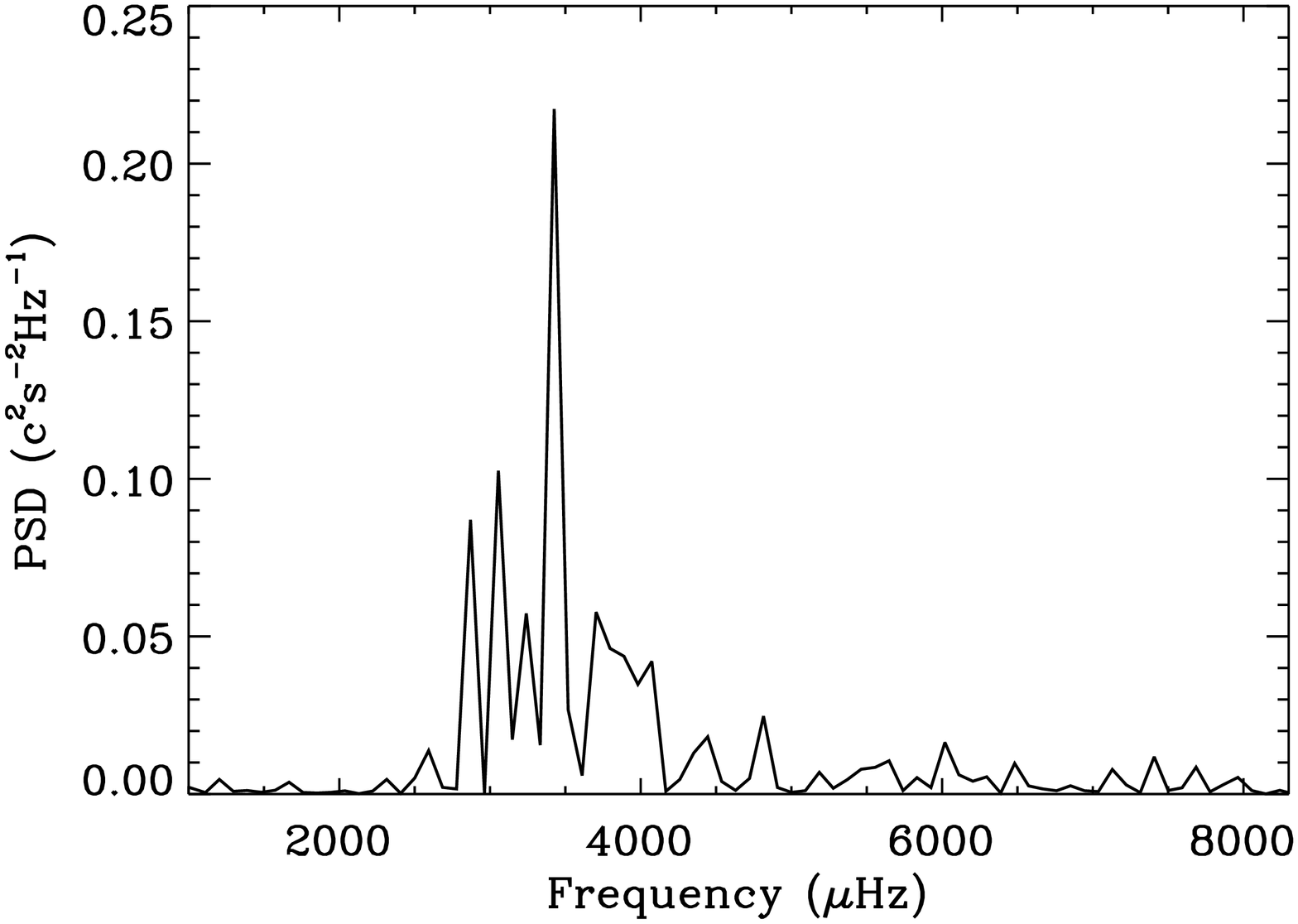}\\
\vspace*{0.25 cm}
\includegraphics[width=0.25\textwidth, angle=90]{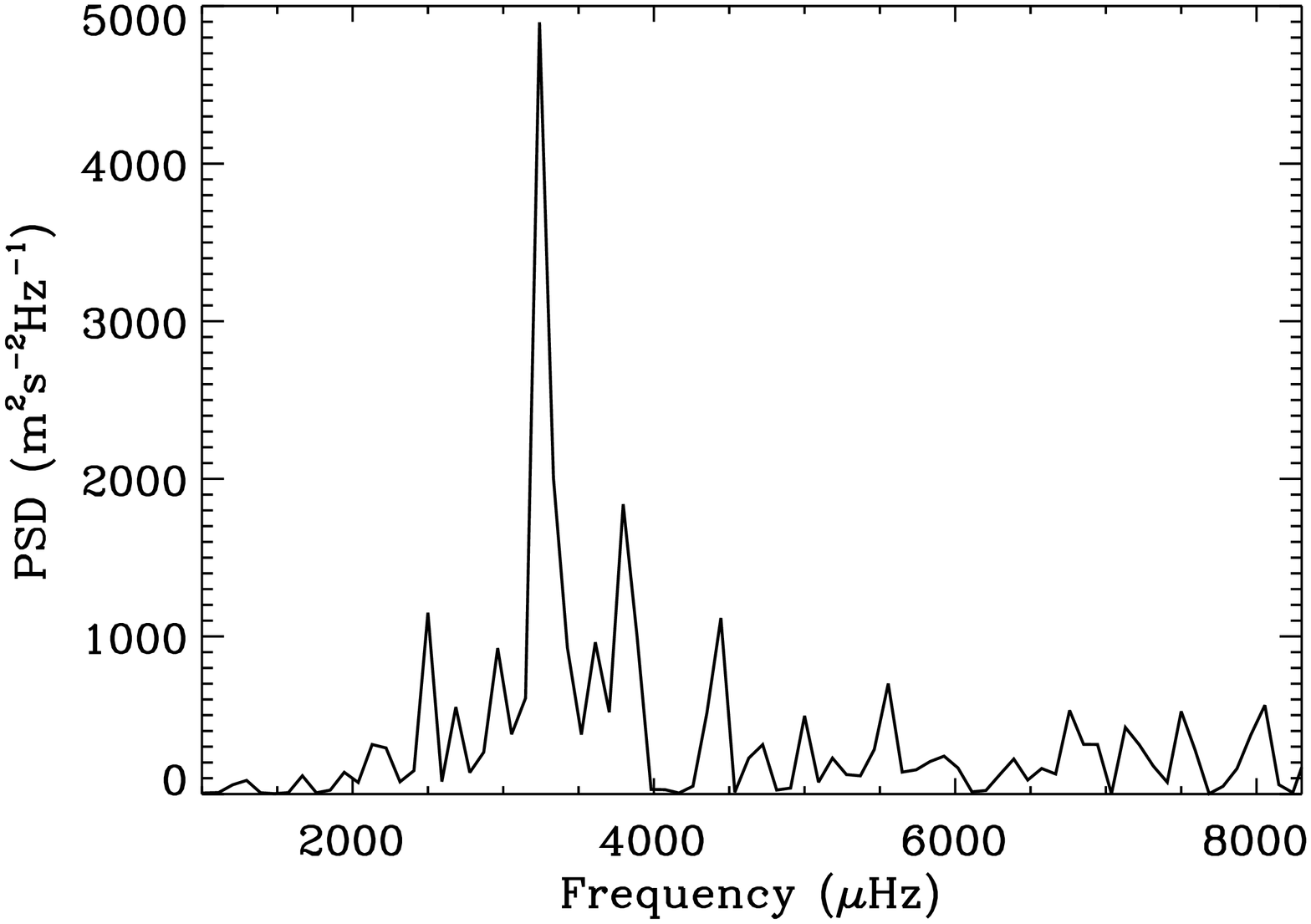} \hspace*{0.25 cm} 
\includegraphics[width=0.25\textwidth, angle=90]{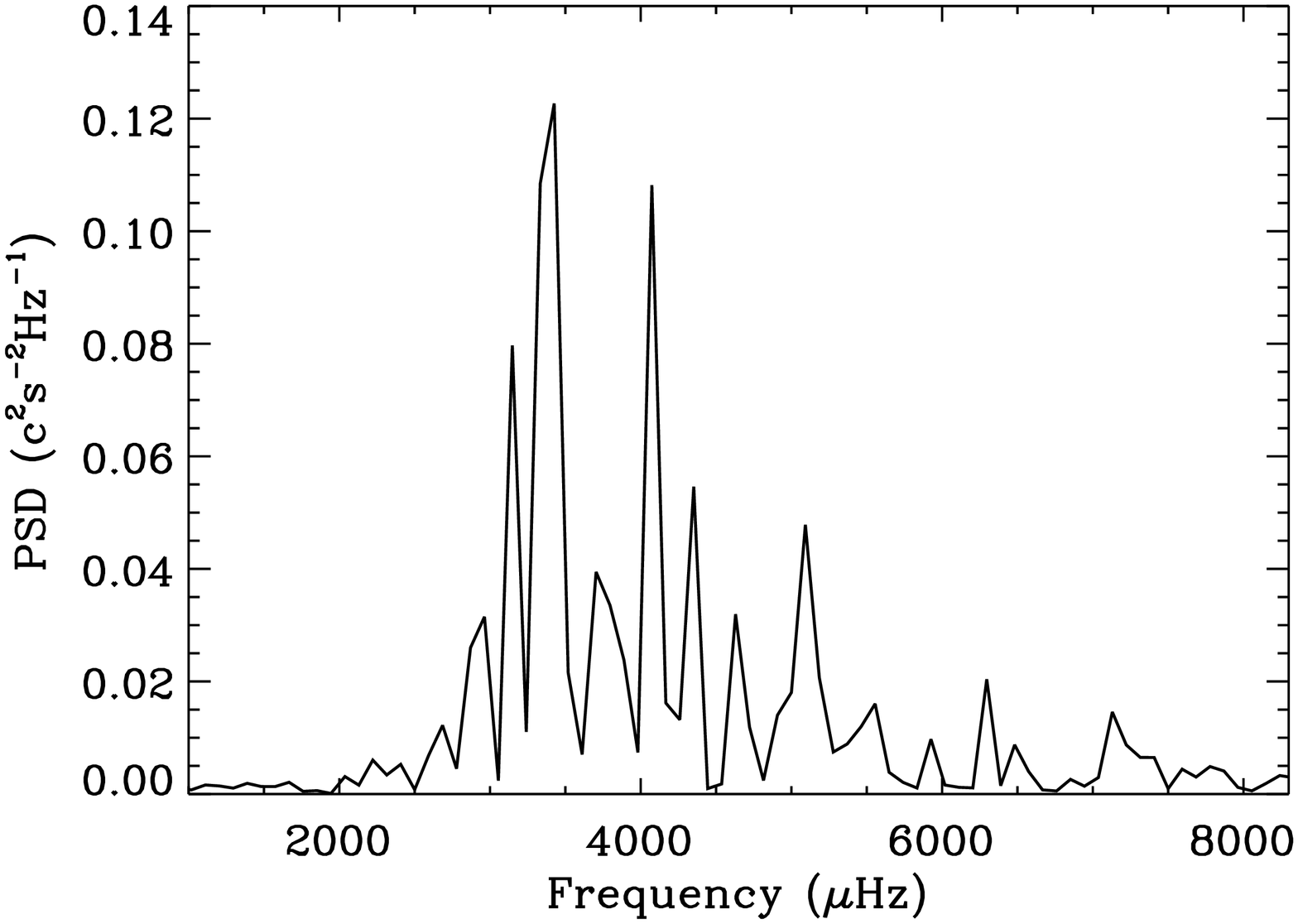}\\
\vspace*{0.25 cm}
\includegraphics[width=0.25\textwidth, angle=90]{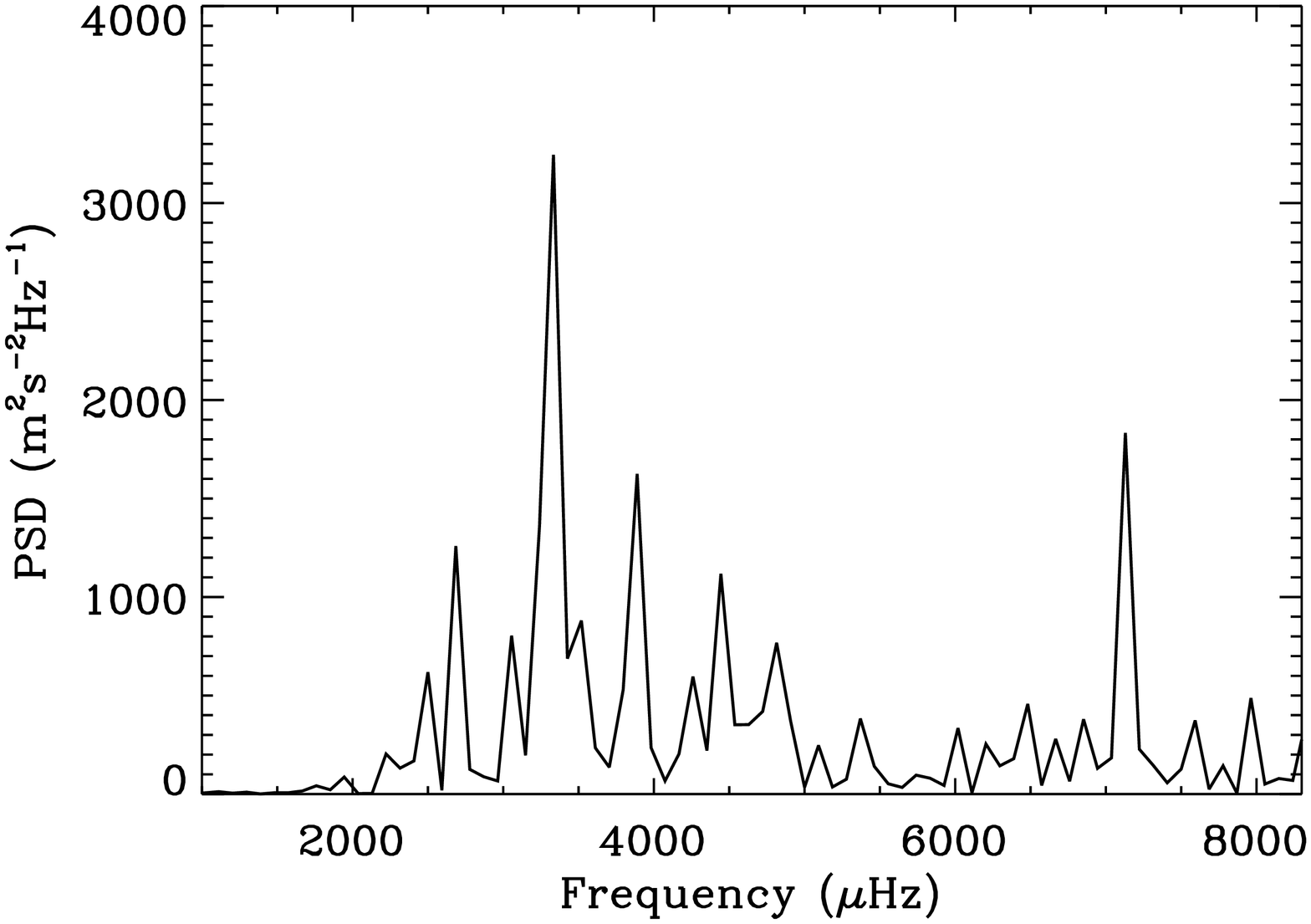} \hspace*{0.25 cm} 
\includegraphics[width=0.25\textwidth, angle=90]{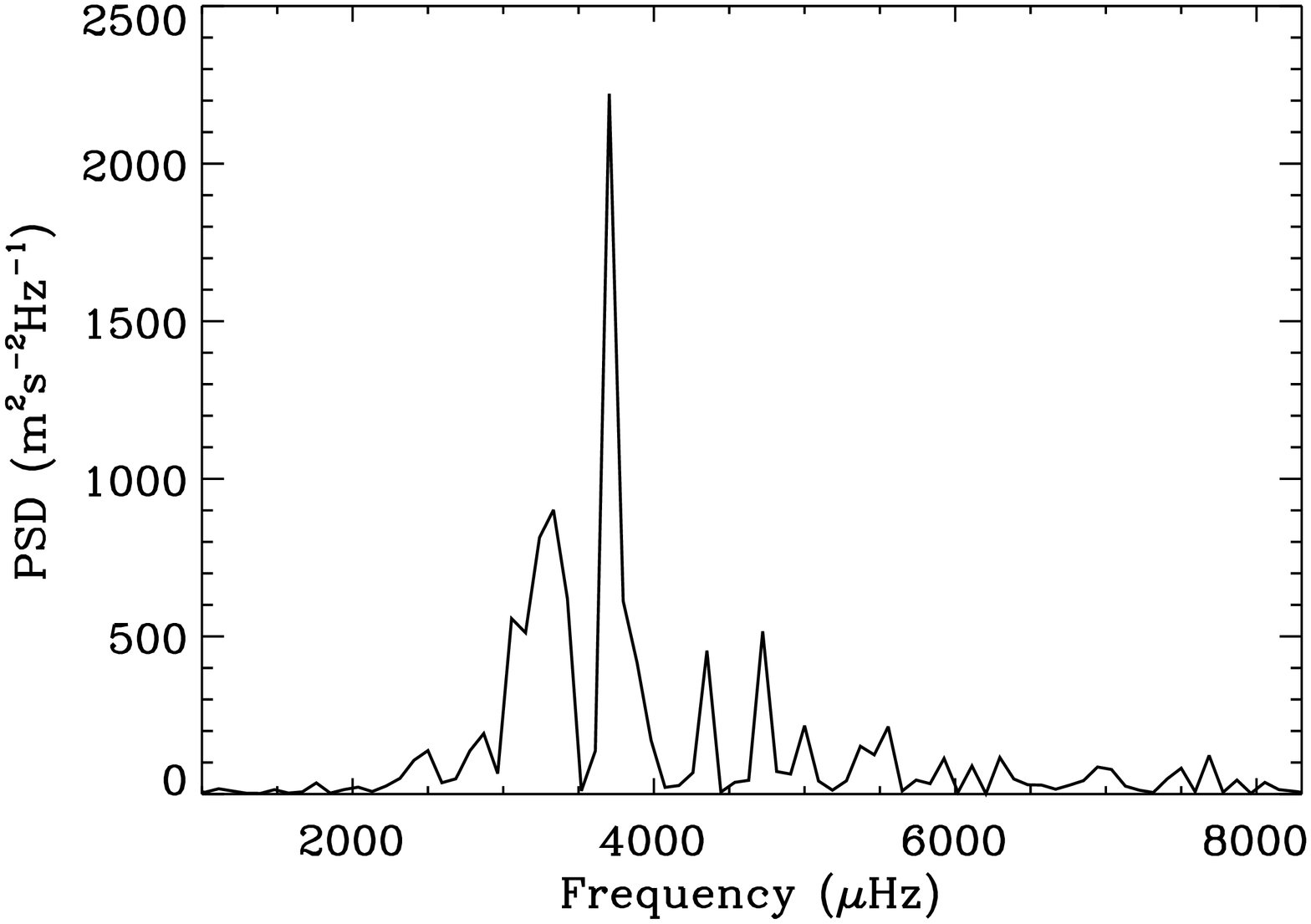}\\
\vspace*{0.25 cm}
\includegraphics[width=0.25\textwidth, angle=90]{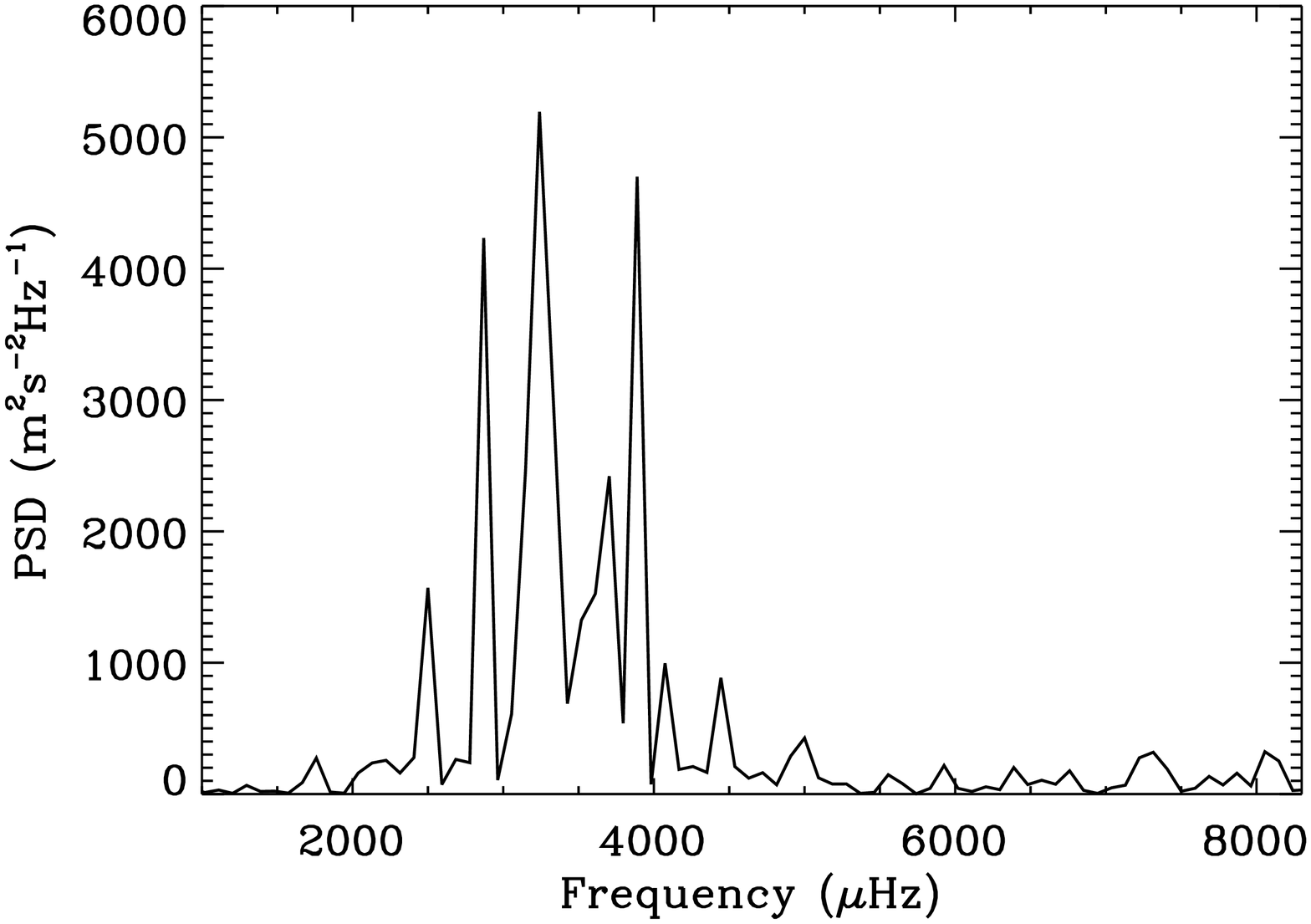} \hspace*{0.25 cm} 
\includegraphics[width=0.25\textwidth, angle=90]{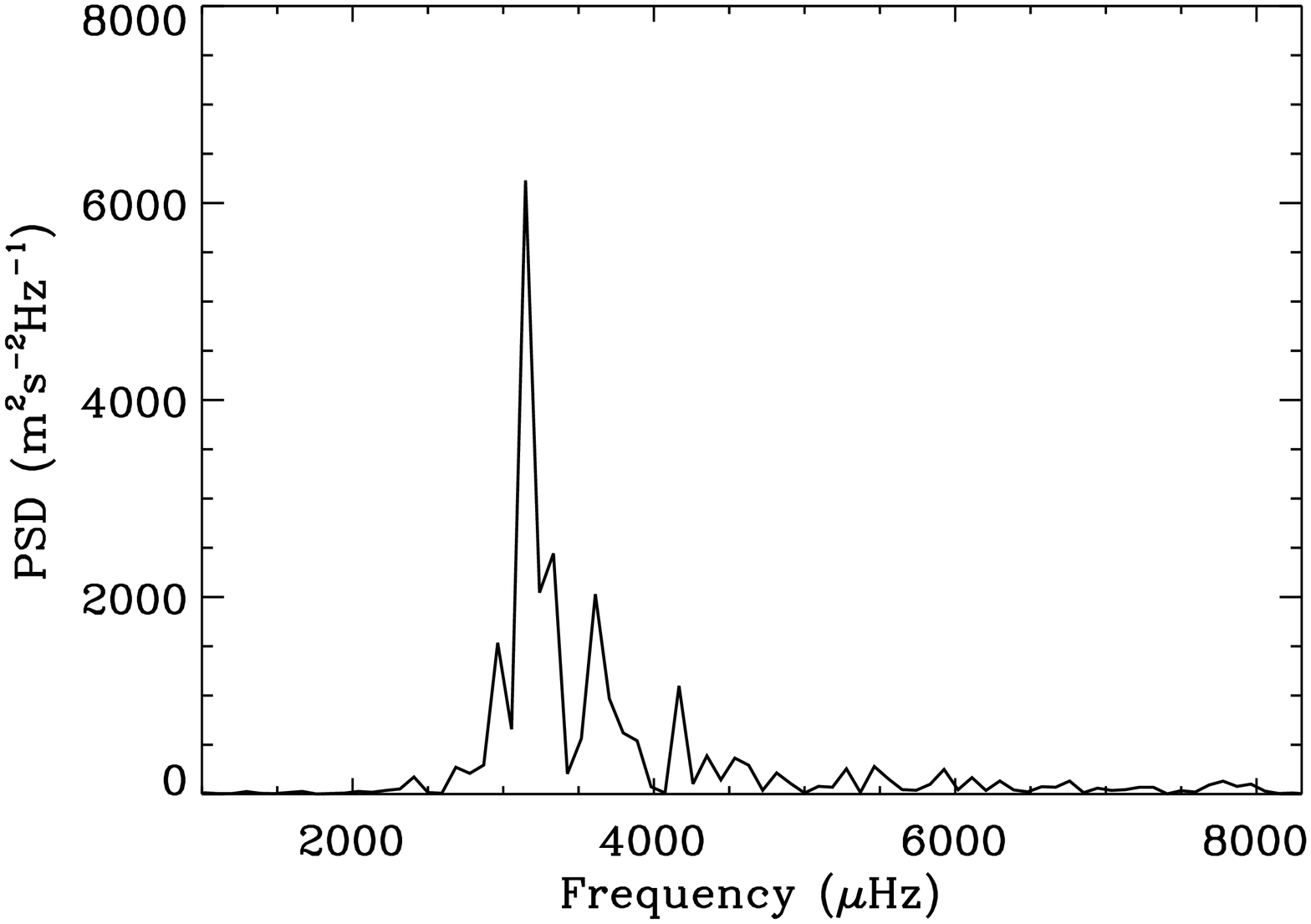}\\
\caption{Fourier Power Spectrum (FPS) estimated from velocity time series. The left panels illustrate the FPS 
obtained from the three hours of the MDI data for the flare events of 28 October 2003 (during 10:00-13:00~UT), 
29 October 2003 (during 19:00-22:00~UT), 6 April 2001 (during 18:00-21:00~UT), and a quiet period (non-flaring 
condition), respectively, from top 
to bottom. The right panels (from top to bottom) show the corresponding FPS 
obtained from the three hours of the GOLF data.}
\end{figure}

\end{document}